\documentstyle[11pt,epsf]{article}

\newlength{\mytopmargin}
\newlength{\myleftmargin}
\def\zz{\relax\hbox{\small \sf Z\kern-.4em Z}}

\begin{document}
\vspace{1cm}
\noindent
\begin{center}{\large \bf  Painlev\'e transcendent
evaluation of the scaled distribution of \\
the smallest eigenvalue in the Laguerre orthogonal and symplectic
ensembles}
\end{center}
\vspace{5mm}
\begin{center}
P.J.~Forrester
\end{center}

\vspace{.2cm}

\noindent
Department of Mathematics and Statistics, University of
Melbourne, Parkville 3052, Australia

\vspace{.2cm}

{\small
\begin{quote} The scaled distribution of the smallest eigenvalue in
the Laguerre orthogonal and symplectic ensembles is evaluated in terms
of a Painlev\'e V transcendent. This same Painlev\'e V transcendent
is known from the work of Tracy and Widom, where it has been shown to specify
the scaled distribution of the smallest eigenvalue
in the Laguerre unitary ensemble. The starting
point for our calculation is the scaled $k$-point distribution of
every odd labelled eigenvalue in two superimposed Laguerre
orthogonal ensembles.
\end{quote}
}
\section{Introduction}
A quite old result in mathematical statistics concerns the eigenvalue
distribution of random matrices of the form $A = X^T X$ where $X$ is
a $n \times N$ ($n \ge N$) rectangular matrix with real entries. First
it was proved by Wishart \cite{Wi28} that with
$$
(dX) := \prod_{j=1}^n \prod_{k=1}^N dx_{jk}, \qquad
(dA) := \prod_{1 \le j < k \le N} da_{jk}
$$
denoting the product of differentials of the independent elements,
the change of variables from the elements of $X$ to the elements
of $A$ takes place according to the formula
\begin{equation}\label{1.1}
(dX) = \Big ( \det A \Big )^{(n-N-1)/2} (dA).
\end{equation}
From this a description in terms of eigenvalues and eigenvectors can be
undertaken by introducing the spectral decomposition
$$
A = O \Lambda O^T
$$
where the columns of $O$ consist of the normalized eigenvectors of $A$, and
$\Lambda$ is the diagonal matrix of eigenvalues. About a decade after
the work 
of Wishart, it was proved by a number of authors (see e.g.~\cite{Hs39}) that
\begin{equation}\label{1.2}
(dA) = \prod_{1 \le j < k \le N} |\lambda_k - \lambda_j|
\prod_{j=1}^N d \lambda_j \, (O^T d O).
\end{equation}
A significant qualitative feature of (\ref{1.2}) is that
the eigenvalue dependence factors from that of the 
eigenvectors.

Suppose now the elements of $X$ are identical, independently distributed
standard Gaussian random variables so that the corresponding probability
measure is proportional to
\begin{equation}\label{1.3}
\prod_{j=1}^n \prod_{k=1}^N e^{-x_{jk}^2/2} (d X) =
e^{- {1 \over 2} {\rm Tr}(X^T X)} (d X) =
e^{- {1 \over 2} \sum_{j=1}^N \lambda_j} (d X).
\end{equation}
Noting that $\det A = \prod_{j=1}^N \lambda_j$, substituting (\ref{1.2})
in (\ref{1.1}) and then substituting the resulting formula for
$(d X)$ in (\ref{1.3}) gives the now standard \cite{Mu82} result that
the eigenvalue probability density function (p.d.f.) of the matrix
$A = X^T X$ is given by
\begin{equation}\label{1.4}
{1 \over C} \prod_{j=1}^N \lambda_j^{(n-N-1)/2} e^{-\lambda_j/2}
\prod_{1 \le j < k \le N} |\lambda_k - \lambda_j|,
\end{equation}
where $C$ denotes the normalization and $\lambda_j > 0$
$(j=1,\dots,N)$. This is referred to as the real Wishart distribution, or
alternatively as the Laguerre orthogonal ensemble (LOE${}_N$), the latter
name originating from the occurence of the classical Laguerre weight
function $\lambda^a e^{-\lambda}$ (up to scaling of $\lambda$)
and the invariance of (\ref{1.4}) under the mapping $A \mapsto O A
O^T$.

Another random matrix structure which leads to the p.d.f.~(\ref{1.4})
is the block matrix
\begin{equation}\label{1.5}
\left ( \begin{array}{cc} 0_{n \times n} & X \\
X^T & 0_{N \times N} \end{array} \right ).
\end{equation}
It is straightforward to verify that in general (\ref{1.5}) has
$n - N$ zero eigenvalues, while the remaining $2N$ eigenvalues come in
$\pm$ pairs. It is similarly easy to verify that with $X$ distributed
according to (\ref{1.3}) the positive eigenvalues are distributed
according to (\ref{1.4}) but with $\lambda_j \mapsto \lambda_j^2$ and
an additional factor of $\prod_{j=1}^N |\lambda_j|$. Hence the square
roots of the positive eigenvalues of (\ref{1.5}) are distributed 
according to (\ref{1.4}).

Over the past decade the p.d.f.~(\ref{1.4}) has found application in a
number of physical problems.  One example is in the theory of quantum
transport through disordered wires, in which the matrix product
$X^T X$ for $X$ a $N \times N$ random matrix modelling the top
right hand block of the transmission matrix occurs in the Landauer 
formula for the conductance (see e.g.~\cite{Be97}). Another is in
quantum chromodynamics, where the structure (\ref{1.5}) models a
random Dirac operator in the chiral guage, and the number of zero
eigenvalues is prescribed \cite{Ve94}.

Because $A$ is positive definite and so has positive eigenvalues the
eigenvalues near $\lambda_j = 0$ are said to be near the hard edge.
For eigenvalues in this neighbourhood, it is known that with $n-N$
fixed, the statistical properties tend to well defined limits in the
$N \to \infty$ scaled limits, where the scaling is
\begin{equation}\label{1.SLa}
\lambda_j \mapsto \lambda_j / 4N
\end{equation}
\cite{Ed88,Fo93a}. In fact the scaled $k$-point distribution function
is known exactly \cite{FNH99}. Thus with
\begin{eqnarray}
a & := & (n-N-1)/2 \nonumber \\
K^{\rm h}(X,Y) & := &
{J_a(X^{1/2})Y^{1/2}J'_a(Y^{1/2}) - X^{1/2}J'_a(X^{1/2})J_a(Y^{1/2}) \over
2(X-Y)} \label{KH} \\
D_1^{\rm h}(x,y) & := & {\partial \over \partial x} S_1(x,y) \nonumber \\
I_1^{\rm h}(x,y) & := &  - \int_x^y S_1(x,z) \, 
dz - {1 \over 2} {\rm sgn}(x-y) \nonumber \\
f_1(x,y) & := &  \left[
\begin{array}{ll}
S_1^{\rm h}(x,y) & I_1^{\rm h}(x,y) \\
D_1^{\rm h}(x,y) & S_1^{\rm h}(y,x)
\end{array} \right]  \nonumber \\
\rho_{(k)}^{\rm LOEh}(x_1,\dots,x_k;a) & := &
\lim_{N \to \infty} \Big ( {1 \over 4N} \Big )^k
\rho_{(k)}^{{\rm LOE}_N} \Big ( {x_1 \over 4N}, \dots, {x_k \over 4N};a \Big )
\end{eqnarray}
we have
\begin{equation}\label{LOE1}
\rho_{(k)}^{\rm LOEh}(x_1,\dots,x_k;(a-1)/2) = {\rm qdet}
\, [f_1(x_j,x_k)]_{j,l=1,\dots,k}
\end{equation}
where qdet is the quaternion determinant introduced into random matrix theory
by Dyson \cite{Dy70}, and the superscipt ``h'' denotes the hard edge
scaling (\ref{1.SLa}).

In this work we will compute the exact distribution of the smallest 
eigenvalue in the scaled LOE as specified by the $k$-point distribution
(\ref{LOE1}) in terms of a certain Painlev\'e V transcendent.
We will also compute the same distribution for the scaled Laguerre symplectic
ensemble (LSE), which before scaling is specified by the 
eigenvalue p.d.f.
$$
{1 \over C} \prod_{j=1}^N \lambda_j^{a} e^{-\lambda_j}
\prod_{1 \le j < k \le N} |\lambda_k - \lambda_j|^4.
$$
This p.d.f.~results from positive definite matrices $A = X^\dagger X$
when the matrix $X$ has the $2 \times 2$ block structure
$$
\left ( \begin{array}{cc} z & w \\
- \bar{w} & \bar{z} \end{array} \right )
$$
of a real quaternion (each eigenvalue is then doubly degenerate as well
as occuring in $\pm$ pairs). The explict form of the corresponding
scaled $k$-point distribution function is given in \cite{FNH99};
in taking the scaled limit with scaling (\ref{1.SLa}) the ensemble
LSE${}_{N/2}$ is considered rather than LSE${}_N$.

Crucial to our study is knowledge, in terms of a Painlev\'e V transcendent,
of the distribution of the smallest eigenvalue in the scaled Laguerre
unitary ensemble (LUE). Before scaling, the latter distribution is
specified by the eigenvalue p.d.f.
\begin{equation}\label{LUEa}
{1 \over C} \prod_{j=1}^N \lambda_j^{a} e^{-\lambda_j}
\prod_{1 \le j < k \le N} |\lambda_k - \lambda_j|^2,
\end{equation}
and results from  positive definite matrices $A = X^\dagger X$
when the matrix $X$ has complex elements. The corresponding
scaled $k$-point distribution function
$$
\rho_{(k)}^{\rm LUEh}(x_1,\dots,x_k) =
\lim_{N \to \infty} \Big ({1 \over 4 N} \Big )^k
\rho_{(k)}^{{\rm LUE}_N} \Big ( {x_1 \over 4N}, \dots, {x_k \over 4N}
\Big )
$$
has the explicit form \cite{Fo93a}
\begin{equation}\label{LUE1}
\rho_{(k)}^{\rm LUEh}(x_1,\dots,x_k) =
\det [K^{\rm h}(x_j,x_l)]_{j,l=1,\dots,k},
\end{equation}
where $K^{\rm h}$ is given by (\ref{KH}).

In general the probability that there are no eigenvalues in an interval
$J$, $E(0;J)$, can be written in terms of the corresponding $k$-point
distribution by
\begin{equation}\label{2a}
E(0;J) = 1 + \sum_{k=1}^\infty {(-1)^k \over k!}
\int_J dx_1 \cdots \int_J dx_k \, \rho_{(k)}(x_1,\dots,x_k).
\end{equation}
For $\rho_{(k)}$ a $k \times k$ determinant with entries $g(x_j,x_k)$ the
structure (\ref{2a}) is just the expansion of the Fredholm integral
operator on $J$ with kernel $g(x,y)$. Thus it follows from (\ref{LUE1})
that 
$$
E_2^{\rm h}(0;(0,s);a) = \det (1 - K^{\rm h})
$$
where $E_2^{\rm h}(0;(0,s);a)$ denotes the probability there are no
eigenvalues in the interval $(0,s)$ for the scaled LUE (the subscript
2 characterizes the LUE via the exponent on the product of differences in
(\ref{LUEa})), while $K^{\rm h}$ denotes the integral operator on
$(0,s)$ with kernel $K^{\rm h}(x,y)$. With $p_\beta^{\rm min}(s;a)$
denoting the distribution of the smallest eigenvalue in the scaled
LOE ($\beta = 1$), LUE ($\beta = 2$) or LSE ($\beta = 4$) we have in
general
$$
p_\beta^{\rm min}(s;a) = - {d \over ds} E_\beta^{\rm h}(0;(0,s);a)
$$
so to compute $p_\beta^{\rm min}(s;a)$ it suffices to compute 
$E_\beta^{\rm h}(0;(0,s);a)$.

For general $a > -1$, $E_2^{\rm h}(0;(0,s);a)$ has been computed in
terms of a Painlev\'e transcendent by Tracy and Widom \cite{TW94b}.
The Painlev\'e transcendent is denoted by $q_{\rm h}$
(in \cite{TW94b} the subscript h is not present), 
and specified as the solution
of the nonlinear equation
\begin{equation}\label{PV}
s(q_{\rm h}^2-1)(sq_{\rm h}')' 
= q_{\rm h}(sq_{\rm h}')^2 + {1 \over 4} (s-a^2)q_{\rm h} +
{1 \over 4} sq_{\rm h}^3(q_{\rm h}^2 - 2)
\end{equation}
subject to the boundary condition
$$
q_{\rm h}(s) \mathop{\sim}\limits_{s \to 0^+} {1 \over 2^a \Gamma(1+a)} s^{a/2}.
$$
That $q_{\rm h}$ is a Painlev\'e transcendent follows from the transformation
 \cite{TW94b}
$$
q_{\rm h}(s) = {1 + y(x) \over 1 - y(x)}, \quad s=x^2
$$
from which one can deduce that $y(x)$ satisfies the Painlev\'e V
equation
$$
 y'' = \Big ( {1 \over 2y} + {1 \over 1 - y} \Big ) (y')^2
- {1 \over x} y' + {(y-1)^2 \over x^2} \Big ( \alpha y + {\beta \over
y} \Big ) + {\gamma y \over x} + {\delta y (y+1) \over y - 1}
$$
with $\alpha = - \beta = a^2/8$, $\gamma = 0$ and $\delta = -2$.
The result of  \cite{TW94b} is that the  Painlev\'e transcendent $q_{\rm h}$
specifies $E_2^{\rm h}$ via the formula
\begin{equation}\label{FFa}
E_2^{\rm h}(0;(0,s);a) = \exp \Big ( - {1 \over 4} \int_0^s
(\log s/t) (q_{\rm h}(t))^2 \, dt \Big ).
\end{equation}

In this work we will show that $E_1^{\rm h}$ can also be evaluated in
terms of $q_{\rm h}(t)$. Specifically, we obtain the formula
\begin{equation}\label{R1}
\Big ( E_1^{\rm h}(0;(0,s);(a-1)/2) \Big )^2 =
E_2^{\rm h}(0;(0,s);a) \exp \Big ( - {1 \over 2}
\int_0^s {q_{\rm h}(t) \over \sqrt{t}} \, dt \Big ).
\end{equation}
With $E_2^{\rm h}$ and $ E_1^{\rm h}$ known in terms of $q_{\rm h}(t)$
the probability $E_4^{\rm h}(0;(0,s);a)$ can also be expressed in terms
of $q_{\rm h}(t)$ by using the formula \cite{FR99}
\begin{equation}\label{1.15'}
E_4^{\rm h}(0;(0,s);a+1) = {1 \over 2} \Big (
 E_1^{\rm h}(0;(0,s);(a-1)/2) + {E_2^{\rm h}(0;(0,s);a) \over
E_1^{\rm h}(0;(0,s);(a-1)/2)} \Big ).
\end{equation}
Thus
\begin{equation}\label{R2}
\Big ( E_4^{\rm h}(0;(0,s);a+1)  \Big )^2 =
E_2^{\rm h}(0;(0,s);a) \cosh^2 \Big ( {1 \over 4}
\int_0^s {q_{\rm h}(t) \over \sqrt{t}} \, dt \Big ).
\end{equation}
Here $E_4^{\rm h}$ is computed by scaling the ensemble LSE${}_{N/2}$
according to (\ref{1.SLa}).
Also, with $ E_1^{\rm h}(1;(0,s);a)$ denoting the probability that there
is exactly one eigenvalue in the interval $(0,s)$ for the scaled LOE, we
have the inter-relationship \cite{FR99}
$$
 E_1^{\rm h}(1;(0,s);(a-1)/2) =
E_4^{\rm h}(0;(0,s);a+1) -  E_1^{\rm h}(0;(0,s);(a-1)/2).
$$
Substituting (\ref{R1}) and (\ref{R2}) shows
\begin{equation}\label{R3}
\Big ( E_1^{\rm h}(1;(0,s);(a-1)/2) \Big )^2 =
E_2^{\rm h}(0;(0,s);a) \sinh^2 \Big ( {1 \over 4}
\int_0^s {q_{\rm h}(t) \over \sqrt{t}} \, dt \Big ).
\end{equation}

Crucial to our derivation of (\ref{R1}) is a reworking of the derivation
of Tracy and Widom \cite{TW96} giving the probability $E_1^{\rm s}
(0;(s,\infty))$ in terms of a Painlev\'e transcendent. Here
$E_1^{\rm s}(0;(s,\infty))$ denotes the probability that there are no
eigenvalues in the interval $(s,\infty)$ for the scaled Gaussian
orthogonal ensemble (GOE). 
As the density falls off rapidly as $s$ increases, the region $(s,\infty)$
is said to be a soft edge, thus explaining the use of the superscript
``s'' in $E_1^{\rm s}$.

The ensemble GOE${}_N$ refers to the eigenvalue
p.d.f.
\begin{equation}\label{2.1}
{1 \over C} \prod_{j=1}^N e^{-x_j^2 / 2} \prod_{1 \le j < k \le N}
|x_k - x_j|.
\end{equation}
This is realized by $N \times N$ real symmetric matrices, with diagonal
elements having the Gaussian distribution $N[0,1]$, and independent off
diagonal elements having the distribution $N[0,1/\sqrt{2}]$. Tracy and
Widom begin with the quaternion determinant expression \cite{Me71} for the 
$k$-point distribution function in the
finite GOE
(we also draw attention to the recent work
\cite{AV99a} in which Painlev\'e type recurrence equations
are obtained for the analogue of the probabilities
$E_1^{\rm s}$ and $E_1^{\rm h}$ in the finite system). Instead, inspired by the observation of Baik and Rains
\cite{BR99a} that the square of the distribution of the largest
eigenvalue in GOE is equal to the distribution of the largest
eigenvalue in two independent, appropriately superimposed GOE's,
technically the ensemble
\begin{equation}\label{DI}
{\rm even}({\rm GOE}_N \cup {\rm GOE}_N),
\end{equation}
we take as our starting point the $k$-point distribution of (\ref{DI}),
scaled at the spectrum edge. 

Now the $k$-point distribution of (\ref{DI}) is an ordinary determinant,
whereas the $k$-point distribution of the GOE is a quaternion determinant.
Furthermore the elements of the determinant contains terms familiar
from the analysis of $E_2^{\rm s}$ given in \cite{TW94a}; this is also true
of the quaternion determinant but the former involves only a subset of the
latter. These facts together provide a simplified evaluation of 
$E_1^{\rm s}$. The power of this derivation is demonstrated by its
application to the evaluation of $E_1^{\rm h}$. We find that each step
used in the derivation of $E_1^{\rm s}$ has an analogous step in the case of
$E_1^{\rm h}$ and this leads to (\ref{R1}).

We begin in Section 2 by providing the evaluation of the $k$-point
distribution for the ensemble (\ref{DI}) in the scaled limit at the
right hand soft edge, as well as that for the ensemble
\begin{equation}\label{DI1}
{\rm odd}( {\rm LOE}_N \cup {\rm LOE}_N )
\end{equation}
in the scaled limit at the hard edge. In Section 3 we begin by using the
evaluation of $\rho_{(k)}$ obtained in Section 2 as the starting point
for the evaluation of $E_1^{\rm s}(0;(s,\infty))$, and then proceed
to mimick this calculation to evaluate $E_1^{\rm h}(0;(0,s);a)$.
In Section 4 our evaluations (\ref{R1}) and (\ref{R2}) are related to
previously known results.

\section{The ensembles even/odd$({\rm OE}_N(f) \cup {\rm OE}_N(f))$ for
$f$ classical}
\setcounter{equation}{0}

Let OE${}_N(f)$ denote the matrix ensemble with eigenvalue p.d.f.
\begin{equation}\label{2.2}
\prod_{j=1}^N f(x_j) \prod_{1 \le j < k \le N} |x_k - x_j|.
\end{equation}
We see from (\ref{1.4}) and (\ref{2.1}) that the LOE is of this form
with 
\begin{equation}\label{2.3}
f(x) = x^a e^{-x/2}, \qquad (x >0, \: a := (n-N-1)/2)
\end{equation}
while the GOE is of this form with
$$
f(x) = e^{-x^2/2}.
$$
In fact the four special choices of $f$
\begin{equation}\label{f}
f(x) = \left \{ \begin{array}{ll}
e^{-x^2/2}, & {\rm Gaussian} \\
x^{(a-1)/2} e^{-x/2} \: (x > 0), & {\rm Laguerre} \\
(1-x)^{(a-1)/2}(1+x)^{(b-1)/2} \: (-1 < x < 1), & {\rm Jacobi} \\
(1+x^2)^{-(\alpha + 1)/2}, & {\rm Cauchy}
\end{array} \right.
\end{equation}
(note that the exponent of $x$ in the Laguerre case has been renormalized
relative to (\ref{2.3})) have been shown in \cite{FR99} to possess special
properties in regards to the superimposed ensembles
\begin{equation}\label{f1}
{\rm even}({\rm OE}_N(f) \cup {\rm OE}_N(f)) \quad {\rm and} \quad
{\rm odd}({\rm OE}_N(f) \cup {\rm OE}_N(f))
\end{equation}
(amongst other superimposed ensembles). In particular the $k$-point
distribution is given by a determinant formula with the same general
structure in each case.

To present the formula for the $k$-point distributions, some additional
theory from \cite{FR99} must be recalled. In particular, it is found that
each of the weight functions (\ref{f}) is one member of a pair
$(f,g)$ which naturally occur in the study of the superimposed ensembles.
Explicitly the weight functions $g$ are
\begin{equation}\label{g}
g = \left \{
\begin{array}{ll} e^{-x^2}, & {\rm Gaussian} \\
x^a e^{-x} \: \: (x >0), & {\rm Laguerre} \\
(1-x)^a(1+x)^b \: \: (-1 < x < 1), & {\rm Jacobi} \\
(1 + x^2)^{-\alpha}, & {\rm Cauchy},
\end{array} \right.
\end{equation}
Now, let $\{p_n(x)\}_{n=0,1,\dots}$ denote the set of monic orthogonal
polynomials associated with a particular weight function $g$, and
let $(p_n,p_n)_2$ denote the corresponding normalization. Then it is
shown in \cite{FR99} that for the ensemble even$({\rm OE}_N(f)
\cup {\rm OE}_N(f))$
\begin{equation}\label{s1}
\rho_{(k)}(x_1,\dots,x_k) =
\prod_{i=1}^k g(x_i) 
\det \bigg [
\sum_{l=0}^{N-2} {p_l(x_i) p_l(x_j) \over (p_l, p_l)_2}
+ {p_{N-1}(x_i) F_{N-1}(x_j) \over (p_{N-1}, F_{N-1})_2}
\bigg ]_{i,j=1,\dots,n},
\end{equation}
where
\begin{equation}\label{s2}
F_{N-1}(x) = \sum_{l=N-1}^\infty {(p_l, I_-)_2 \over (p_l,p_l)_2} p_l(x),
\quad I_-(x) := {f(x) \over g(x)} 
\int^x_{-\infty} f(t) \, dt.
\end{equation}
Similarly for the ensemble odd$({\rm OE}_N(f) \cup {\rm OE}_N(f))$ the
$k$-point distribution is again given by the formula (\ref{s1}) but with
$I_-$ in (\ref{s2}) replaced by
\begin{equation}\label{s3}
I_+(x) := {f(x) \over g(x)} \int_x^\infty f(t) \, dt.
\end{equation}

The summation in (\ref{s1}) can be evaluated according to the 
Christoffel-Darboux formula, and the corresponding scaling limits are well
known \cite{Fo93a}. To compute the scaled limit of the quantity $F_{N-1}$
in (\ref{s1}), we first make use of results from the work \cite{AFNV99}
to provide the explicit evaluation of the coefficients
$$
(p_l,I_-)_2 := \int_{-\infty}^\infty dx \, f(x) p_l(x)
\int_{-\infty}^x dt \, f(t),
$$
applicable in all the classical cases (\ref{f}). First we note
\begin{equation}\label{cp}
\int_{-\infty}^x f(t) \, dt = {1 \over 2} \int_{-\infty}^\infty
{\rm sgn} (x-t) f(t) \, dt + {1 \over 2} \int_{-\infty}^\infty f(t) \,
dt =: \tilde{\phi}_0(x) + \tilde{s}_0,
\end{equation}
which allows us to write
\begin{equation}\label{ftt}
(p_l,I_-)_2 = \int_{-\infty}^\infty f(x) p_l(x)  \tilde{\phi}_0(x) \, dx
+  \tilde{s}_0 \int_{-\infty}^\infty f(x) p_l(x) \, dx.
\end{equation} 
Now results in \cite{AFNV99} give
$$
 \tilde{\phi}_0(x) = {1 \over \gamma_0} {g(x) \over f(x)}
\sum_{\nu = 0}^\infty \prod_{k=1}^\nu
\Big ({\gamma_{2k-1} \over  \gamma_{2k}} \Big ) {p_{2\nu + 1}(x) \over
(p_{2\nu+1}, p_{2\nu+1})_2}
$$
where
\begin{equation}\label{ga}
\gamma_k (p_k,p_k)_2 = \left \{
\begin{array}{ll} 1, & {\rm Gaussian}\\[.2cm] {1 \over 2}, & {\rm Laguerre} 
\\[.2cm] {1 \over 2}(2k+2+a+b), & {\rm Jacobi} \\[.2cm]
\alpha - 1 - k, & {\rm Cauchy}  \end{array} \right.
\end{equation}
This allows us to immediately evaluate the first term in (\ref{ftt}).

It remains to compute the second term in (\ref{ftt}). With the notation
$$
 \tilde{s}_l := {1 \over 2} \int_{-\infty}^\infty f(x) p_l(x) \, dx.
$$
this term is given by $2  \tilde{s}_0 \tilde{s}_l$.
Consider first the case $l$ even. With
$$
\tilde{\phi}_l(x) := {1 \over 2} \int_{-\infty}^\infty {\rm sgn}(x-t)
p_l(t) f(t) \, dt
$$
we know from \cite{AFNV99} that
$$
\tilde{\phi}_{2k}(x) = {1 \over \gamma_{2k}}
{1 \over \prod_{l=1}^k (\gamma_{2l-1} / \gamma_{2l})}
{g(x) \over f(x)} \sum_{\nu = k}^\infty 
\prod_{l=1}^\nu \Big ( {\gamma_{2l-1} \over \gamma_{2l}} \Big )
{p_{2\nu + 1}(x) \over (p_{2\nu +1}, p_{2\nu+1})_2}.
$$
Forming the ratio $\tilde{\phi}_{2k}(x) / \tilde{\phi}_{2k-2}(x)$
and taking the limit $x \to \infty$ shows that
$\tilde{s}_{2k}/ \tilde{s}_{2k-2} = \gamma_{2k-2} / \gamma_{2k-1}$ and
thus we have the evaluation
$$
\tilde{s}_{2l} = \tilde{s}_0 \prod_{j=0}^{l-1} {\gamma_{2j} \over
\gamma_{2j+1}}.
$$
To evaluate $\tilde{s}_l$, $l$ odd, we recall the formula
 \cite{AFNV99}
$$
\tilde{\phi}_{2k+1}(x) - {\gamma_{2k-1} \over \gamma_{2k}}
\tilde{\phi}_{2k-1}(x) = - {1 \over \gamma_{2k}} {g(x) \over f(x)}
{p_{2k}(x) \over (p_{2k},p_{2k})_2}.
$$
Taking the limit $x \to \infty$ implies $\tilde{s}_{2k+1} =
(\gamma_{2k-1}/\gamma_{2k}) \tilde{s}_{2k - 1}$ and since
$\tilde{s}_{-1} := 0$ this gives
$$
\tilde{s}_{2l+1} = 0.
$$
Thus the second term in (\ref{ftt}) is fully determined.

Substituting the evaluation of $(p_l,I_-)_2$ obtained from the above working
in (\ref{s2}) shows
\begin{equation}\label{FF}
F_{N-1}(x) =  {1 \over \gamma_0}
\sum_{\nu = [(N-1)/2]}^\infty
{\prod_{l=1}^\nu (\gamma_{2l-1} / \gamma_{2l})
 \over (p_{2 \nu + 1}, p_{2 \nu + 1})_2}
p_{2 \nu + 1}(x)
+ 2\tilde{s}_0^2
 \sum_{l=[N/2]}^\infty {\prod_{j=0}^{l-1}(\gamma_{2j} / \gamma_{2j+1})
 \over (p_{2l}
, p_{2l})_2} p_{2l}(x).
\end{equation}
From this result we read off that
\begin{equation}\label{FF1}
(p_{N-1}, F_{N-1})_2 =
\left \{ \begin{array}{ll} {1 \over \gamma_0}
\prod_{l=1}^{(N-2)/2} (\gamma_{2l-1} / \gamma_{2l}) & N \: \: {\rm even}
\\[.2cm]
{2\tilde{s}_0^2}
  \prod_{j=1}^{(N-1)/2}
(\gamma_{2j-2} / \gamma_{2j-1})
 & N \: \: {\rm odd} \end{array} \right.
\end{equation}
so all quantities in the expression (\ref{s1}) for $\rho_{(k)}$ are now
known explicitly.

In the case of the ensemble odd$({\rm OE}_N(f) \cup {\rm OE}_N(f))$ the
definition (\ref{s2}) of $F_{N-1}$ has $I_-$ replaced by $I_+$. Noting
$$
\int_x^\infty f(t) \, dt = - {1 \over 2} \int_{-\infty}^\infty
{\rm sgn}(x-t) f(t) \, dt + {1 \over 2} \int_{-\infty}^\infty f(t) \, dt
= - \tilde{\phi}_0(x) + \tilde{s}_0
$$
which differs from (\ref{cp}) only in the sign of the first term, we see by
revising the working which led from (\ref{cp}) to (\ref{FF}) the only
modification needed to the formula (\ref{FF}) is that a minus sign be
placed in front of the first term (and similarly in (\ref{FF1})).

\subsection{Guassian ensemble at the soft edge}
In the Gaussian case
\begin{equation}\label{go}
f(x) = e^{-x^2/2}, \quad g(x) = e^{-x^2}, \quad
p_l(x) = 2^{-l} H_l(x), \quad 
(p_l,p_l)_2 = \pi^{1/2} 2^{-l} l!,
\end{equation}
where $H_l(x)$ denotes the Hermite polynomial.
The soft edge scaling is \cite{Fo93a}
\begin{equation}\label{sc}
x = (2N)^{1/2} + {X \over 2^{1/2} N^{1/6}}
\end{equation}
so we seek to compute
$$
\rho_{(k)}^{{\rm (GOE)}^2 {\rm s}}(X_1,\dots,X_k) :=
\lim_{N \to \infty} \Big ( {1 \over 2^{1/2} N^{1/6}} \Big )^k
\rho_{(k)}^{\rm E}\Big ( (2N)^{1/2} + {X_1 \over 2^{1/2} N^{1/6}}, \dots,
(2N)^{1/2} + {X_k \over 2^{1/2} N^{1/6}} \Big )
$$ 
where E denotes the ensemble (\ref{DI}) and the r.h.s.~is given by
(\ref{s1}) with the substitutions (\ref{go}).

Regarding the summation in (\ref{s1}), we know from the study of the GUE at
the soft edge that \cite{Fo93a}
$$
\lim_{N \to \infty} {1 \over 2^{1/2} N^{1/6}}
\sum_{l=0}^{N-2} {p_l(x) p_l(y) \over (p_l,p_l)_2}
\Big |_{x = (2N)^{1/2} + X/2^{1/2} N^{1/6} \atop
y = (2N)^{1/2} + Y/2^{1/2} N^{1/6}} = K^{\rm s}(X,Y),
$$
where, with ${\rm Ai}(x)$ denoting the Airy function
\begin{equation}\label{krs}
K^{\rm s}(x,y) : = { {\rm Ai}\, (x) {\rm Ai}'(y) - {\rm Ai}'(x)
{\rm Ai}\, (y)
\over x - y}.
\end{equation}
This is obtained using the
Christoffel-Darboux summation formula and the asymptotic expansion
\cite{Sz75}
\begin{equation}\label{asy}
e^{-x^2/2} H_n(x) = \pi^{-3/4} 2^{n/2 + 1/4} (n!)^{1/2} n^{-1/12}
\Big ( \pi {\rm Ai}\,(-u) + {\rm O}(n^{-2/3}) \Big )
\end{equation}
where $x = (2n)^{1/2} - u/2^{1/2} n^{1/6}$.

It remains to compute the scaled limit of the term involving
$F_{N-1}$ in (\ref{s1}). Now substituting the values of $\gamma_k$ and
$(p_k,p_k)_2$ from (\ref{ga}) and (\ref{go}), and noting
$\tilde{s}_0^2 = \pi / 2$, (\ref{FF}) reads
\begin{equation}\label{m2}
F_{N-1}(x) =  \pi^{1/2} \sum_{\nu = [(N-1)/2]}^\infty
{\nu! \over (2 \nu + 1)!} H_{2\nu + 1}(x) +
\pi \sum_{l=[N/2]}^\infty {1 \over 2^{2l} l!} H_{2l}(x).
\end{equation}
Next we want to combine this result with (\ref{FF1}). For definiteness
take $N$ to be even. Then we see that
\begin{equation}\label{m3}
\Big ( g(x) g(y) \Big )^{1/2}
 {p_{N-1}(x) F_{N-1}(y) \over (p_{N-1},F_{N-1})_2}
= e^{-x^2/2} 2^{-(N-1)} H_{N-1}(x) \Big ( A_1(y) + A_2(y) \Big )
\end{equation}
with
\begin{eqnarray}\label{2.39'}
A_1(y) & := & {e^{-y^2/2} \over (N/2 - 1)!} \sum_{\nu = 0}^\infty
{(N/2 - 1 + \nu)! \over (N-1+2\nu)!} H_{N-1+2\nu}(y), \nonumber \\
A_2(y) & :=  &  {\pi^{1/2} e^{-y^2/2} \over (N/2 - 1)!} \sum_{l=0}^\infty
{1 \over 2^{N+2l}(N/2 + l)!} H_{N+2l}(y).
\end{eqnarray}
We remark that
the summation defining $A_1(y)$ (with the lower terminal $\nu=0$ replaced
by $\nu=1$) occurs in the study of the soft edge distribution at $\beta = 1$
\cite{FNH99}, and furthermore a procedure has been given to compute its
asymptotic behaviour with the scaling (\ref{sc}), the key ingredient
of which is the asymptotic expansion (\ref{asy}).

For the $x$-dependent terms in (\ref{m3}), (\ref{asy}) gives
$$
e^{-x^2/2} 2^{-(N-1)} H_{N-1}(x) \Big |_{x = (2N)^{1/2} + X/2^{1/2} N^{1/6}}
\: \sim \: \pi^{1/4} 2^{-(N-1)/2 + 1/4} (N-1)!^{1/2} N^{-1/12} {\rm Ai}(X).
$$
For $A_1(y)$, first note that for large $N$
$$
{(N/2 - 1 + \nu)! (N-1)!^{1/2} \over
(N-1+2\nu)!^{1/2} (N/2-1)!} \sim 2^{-\nu}.
$$
Then use of (\ref{asy}) with $n=N+2\nu-1$, $-u \sim Y - 2 \nu / N^{1/3}$ shows
that the sum becomes the Riemann approximation to a definite integral, and
thus
$$
(N-1)!^{1/2}A_1((2N)^{1/2} + Y/ 2^{1/2} N^{1/6}) \sim
2^{(N-1)/2} 2^{1/4} \pi^{-1/4} N^{-1/12} {N^{1/3} \over 2}
\int_0^\infty {\rm Ai}(Y-v) \, dv.
$$
Hence
\begin{equation}\label{c1}
\lim_{N \to \infty} {1 \over 2^{1/2} N^{1/6}}
e^{-x^2 / 2} 2^{-(N-1)} H_{N-1}(x) A_1(y) \Big |_{
x = (2N)^{1/2} + X/2^{1/2} N^{1/6} \atop
y = (2N)^{1/2} + Y/2^{1/2} N^{1/6}}
= {1 \over 2} {\rm Ai}(X) \int_0^\infty {\rm Ai}(Y - v) \, dv.
\end{equation}
Similarly, for $A_2(y)$, noting that for large $N$
$$
{(N+2l)!^{1/2} \over 2^{N+2l} (N/2 + l)!}
{(N-1)!^{1/2} \over (N/2 - 1)!} \sim {1 \over (2 \pi)^{1/2}} 2^{-l},
$$
and using (\ref{asy}) with $n=N+2 \nu$, $-u \sim Y - 2 \nu / N^{1/3}$ we find
$$
(N-1)!^{1/2} A_2((2N)^{1/2} + Y/ 2^{1/2} N^{1/6}) \sim
{1 \over \pi^{1/2}} A_1((2N)^{1/2} + Y/ 2^{1/2} N^{1/6})
$$
and hence
\begin{equation}\label{c2}
\lim_{N \to \infty} {1 \over 2^{1/2} N^{1/6}}
e^{-x^2 / 2} 2^{-(N-1)} H_{N-1}(x) A_2(y) \Big |_{
x = (2N)^{1/2} + X/2^{1/2} N^{1/6} \atop
y = (2N)^{1/2} + Y/2^{1/2} N^{1/6}}
=  {1 \over 2} {\rm Ai}(X) \int_0^\infty {\rm Ai}(Y - v) \, dv.
\end{equation}
The contributions (\ref{c1}) and (\ref{c2}) thus reinforce, so after
adding to (\ref{krs}) we obtain
\begin{equation}\label{c3}
\rho_{(k)}^{{\rm (GOE)}^2 {\rm s}}(X_1,\dots,X_k) =
\det \Big [ \Big (
K^{\rm s}(X_j,X_l)
+ {\rm Ai}\,(X_j) \int_0^\infty {\rm Ai}(X_l - v) \, dv \Big )
\Big ]_{j,l=1,\dots,k}.
\end{equation}

\subsection{Laguerre ensemble at hard edge}
In the Laguerre case
$$
f(x) = x^{(a-1)/2} e^{-x/2}, \qquad g(x) = x^a e^{-x}, \: \: (x >0)
$$
\begin{equation}\label{3.5}
p_l(x) = (-1)^l l! L_l^a(x), \qquad (p_l,p_l)_2 = \Gamma(l+1)
\Gamma(a+l+1),
\end{equation}
where $L_l^a$ denotes the Laguerre polynomial.
At the hard edge the appropriate scaling is specified by (\ref{1.SLa}), so
the task is to compute
\begin{equation}\label{4.50}
\rho_{(k)}^{{\rm (LOE)}^2 {\rm h}}(X_1,\dots,X_k;a) :=
\lim_{N \to \infty} \Big ( {1 \over 4N} \Big )^k
\rho_{(k)}^{\rm E} \Big ( {X_1 \over 4N}, \dots, {X_k \over 4N} \Big )
\end{equation}
where E denotes the ensemble (\ref{DI1}) and $\rho_{(k)}$ on the
r.h.s.~is specified by (\ref{s1}).

The scaled limit of the summation in (\ref{s1}) at the hard edge occurs in
the study of the LUE and is known \cite{Fo93a}. Thus making use of the
Christoffel-Darboux formula and the large $n$ asymptotic formula
\cite{Sz75}
\begin{equation}\label{4.52}
 e^{-y/2} y^{a/2} L_{n}^a(y) \sim n^{a/2} J_a(2\sqrt{ny}).
\end{equation}
one finds
\begin{equation}\label{2.45'}
\lim_{N \to \infty} {1 \over 4N} \sum_{l=0}^{N-2}
{p_l(x) p_l(y) \over (p_l,p_l)_2} \bigg |_{x = X/4N \atop y=Y/4N}
= K^{\rm h}(X,Y)
\end{equation}
where $ K^{\rm h}$ is specified by (\ref{KH}).

The first step in computing the term involving $F_{N-1}$ in (\ref{s1})
is to substitute the formulas (\ref{3.5}) in (\ref{FF}) modified so that
there is a minus sign before the first term (recall the paragraph below
(\ref{FF1}). This gives
\begin{eqnarray*}
F_{N-1}(x) & = &  {2 a! \over (a/2)!} \sum_{\nu = [(N-1)/2]}^\infty
{2^{2\nu} (a/2 + \nu)! \nu! \over (a+2\nu + 1)!} L_{2\nu+1}^a(x) \\
&& + 2^a{((a-1)/2)!^2 (a/2)! \over a!}
\sum_{l=[N/2]}^\infty {(2l)! \over 2^{2l} l! (a/2 + l)!} L_{2l}^a(x).
\end{eqnarray*}
We see from this formula and (\ref{FF1}) (for definiteness in the latter
we take $N$ to be even; recall that in this case a minus sign must be
inserted) that
\begin{eqnarray}\label{f6}\lefteqn{
\Big ( g(x)  g(y) \Big )^{1/2}
 {p_{N-1}(x) F_{N-1}(y) \over (p_{N-1}, F_{N-1})_2}
 =  (g (x))^{1/2} L_{N-1}^a(x) \Big (
{(N-1)! \over 2^{N-2} ((N-2)/2)! (a/2 + (N-2)/2)!} \Big )} \nonumber \\
&& \times \bigg ( \sum_{\nu = (N-2)/2}^\infty
{2^{2\nu} (a/2 + \nu)! \nu! \over (a + 2\nu + 1)!} (g(y))^{1/2} 
L_{2\nu+1}^a(y) \nonumber \\
&& + 2^{a-1} {((a-1)/2)!^2 (a/2)!^2 \over a!^2}
\sum_{l=N/2}^\infty {(2l)! \over 2^{2l} l! (a/2 + l)!} (g(y))^{1/2}
 L_{2l}^a(y)
\bigg ) \nonumber \\
& := &  (g(x))^{1/2} L_{N-1}^a(x)
{(N-1)! \over 2^{N-2} ((N-2)/2)! (a/2 + (N-2)/2)!}
\Big ( B_1(y) + B_2(y) \Big ),
\end{eqnarray}
where $B_1$ and $B_2$ denote the two terms in the final brackets of the line
before.

Consider the function $B_1(y)$. From Stirling's formula we have
$$
{2^{2\nu} (a/2 + \nu)! \nu! \over (a + 2\nu + 1)!} \sim
\pi^{1/2} 2^{-(a+1)} \nu^{-a/2-1/2}.
$$
Using this and (\ref{4.52}) we that
the sum defining $B_1(y)$ is
the Riemann approximation to a definite integral and we find
\begin{equation}\label{he6}
B_1\Big ({Y \over 4N} \Big ) \sim
{\pi^{1/2} \over 2^{a/2 + 1}} \Big ({N \over 2} \Big )^{1/2}
\int_1^\infty t^{-1/2} J_a(\sqrt{tY}) \, dt
= {\pi^{1/2} \over 2^{a/2}} \Big ({N \over 2} \Big )^{1/2}
{1 \over Y^{1/2}} \int_{Y^{1/2}}^\infty J_a(t) \, dt.
\end{equation}
Another application of  Stirling's formula shows
$$
{(N-1)! \over 2^{N-2} ((N-2)/2)! (a/2 + (N-2)/2)!} \sim
2^{(a+1)/2} \pi^{-1/2} N^{(1-a)/2}
$$
so we see from further use of (\ref{4.52}) together with (\ref{he6}) that
\begin{eqnarray}\label{he7}
&\lim_{N \to \infty}
(g(x))^{1/2} {1 \over 4N} L_{N-1}^a(x)
{(N-1)! \over 2^{N-2} ((N-2)/2)! (a/2 + (N-2)/2)!}   B_1(y)
\Big |_{x = X/4N \atop y = Y/4N}& \nonumber \\ &
= {J_a(\sqrt{X}) \over 4 \sqrt{Y}}
\int_{Y^{1/2}}^\infty J_a(t) \, dt.&
\end{eqnarray}
A similar analysis applied to $B_2(y)$ shows
\begin{eqnarray}\label{he8}
&\lim_{N \to \infty} {1 \over 4N}
(g(x))^{1/2} L_{N-1}^a(x)
{(N-1)! \over 2^{N-2} ((N-2)/2)! (a/2 + (N-2)/2)!}  B_2(y)
\Big |_{x = X/4N \atop y = Y/4N}& \nonumber \\ &
= 
{J_a(\sqrt{X}) \over 4 \sqrt{Y}}
 \int_{Y^{1/2}}^\infty  J_a(t) \, dt,&
\end{eqnarray}
which thus reinforces (\ref{he7}). Thus adding twice (\ref{he7}) to
(\ref{2.45'}) we have
\begin{equation}\label{he10}
\rho_{(k)}^{({\rm LOE})^2 \rm h}(X_1,\dots,X_k;(a-1)/2) =
\det \Big [\Big ( K^{\rm h}(X_j,X_l)
+ {J_a(\sqrt{X_j}) \over 2 \sqrt{X_l}}
\int_{\sqrt{X_l}}^\infty  J_a(t) \, dt \Big )
\Big ]_{j,l=1,\dots,k}.
\end{equation}

\section{Gap probabilities at the spectrum edge}
\setcounter{equation}{0}

\subsection{The probability $E_1^{\rm s}(0;(s,\infty))$}
It was remarked in the Introduction that the probability
 $E_1^{\rm s}(0;(s,\infty))$ has been computed in terms of a
Painlev\'e II transcendent by Tracy and Widom \cite{TW96}. Expliclity
let $q_{\rm s}$ denote the solution of the particular Painlev\'e II
differential equation
\begin{equation}\label{2.53}
(q_{\rm s})'' = sq_{\rm s} + 2(q_{\rm s})^3
\end{equation}
subject to the boundary condition $q_{\rm s}(s) \sim {\rm Ai}(s)$ as
$s \to \infty$. Then it is shown in \cite{TW96} that
\begin{equation}\label{TW1}
\Big ( E_1^{\rm s}(0;(s,\infty)) \Big )^2  =  E_2^{\rm s}(0;(s,\infty))
\exp \Big ( - \int_s^\infty q_{\rm s}(t) \, dt \Big )
\end{equation}
where \cite{TW94a}
\begin{equation}\label{3.53'}
E_2^{\rm s}(0;(s,\infty))  =  \exp \Big ( - \int_s^\infty (t-s) 
q_{\rm s}^2(t) \, dt \Big ).
\end{equation}
Here we will use the evaluation of the $k$-point distribution (\ref{c3})
to provide a simplified derivation of (\ref{TW1}) while still following
the essential strategy of \cite{TW96}.

By definition of (\ref{DI}) it follows that
$$
\Big ( E_1^{{\rm GOE}_N}(0;(s,\infty)) \Big )^2 =
E^{{\rm even}({\rm GOE}_N \cup {\rm GOE}_N)}(0;(s,\infty)),
$$
which with the scaling (\ref{sc}) implies
\begin{equation}\label{A1}
\Big ( E_1^{\rm s}(0;(s,\infty)) \Big )^2 = E^{{\rm (GOE)}^2
\rm s}(
0;(s,\infty)).
\end{equation}
Now, recalling the determinant formula (\ref{c3}), we see from (\ref{2a})
and the text immediately below that $ E^{{\rm (GOE)}^2s}$ can be
written as the determinant of a Fredholm integral operator. Thus
\begin{equation}\label{8.stt}
\Big ( E_1^{\rm s}(0;(s,\infty)) \Big )^2 =
\det \Big ( 1 - (K^{\rm s} + A \otimes B) \Big )
\end{equation}
where $K^{\rm s}$ is the integral operator on $(s,\infty)$ with
kernel (\ref{krs}) while $A$ is the operator which multiplies by
${\rm Ai}(x)$, while $B$ is the integral operator with kernel
$\int_0^\infty {\rm Ai}(y-v) \, dv$.

Removing $(1 - K^{\rm s})$ as a factor from (\ref{8.stt}) and recalling
\cite{TW94a}
$$
\det (1 - K^{\rm s}) = E_2^{\rm s}(0;(s,\infty))
$$
we obtain
\begin{eqnarray}\label{8.stt1}
\Big ( E_1^{\rm s}(0;(s,\infty)) \Big )^2 & = &
E_2^{\rm s}(0;(s,\infty)) \det \Big (
1 - (1 - K^{\rm s})^{-1} A \otimes B \Big ) \nonumber \\
& = & E_2^{\rm s}(0;(s,\infty))
\Big ( 1 - \int_s^\infty (1 - K^{\rm s})^{-1} A[y] B(y) \, dy \Big )
\end{eqnarray}
where the second equality follows from the fact that
$(1 - K^{\rm s})^{-1} A[y]$ is the eigenfunction of the operator
$(1 - K^{\rm s})^{-1} A \otimes B$, so the eigenvalue is
$$
 \int_s^\infty (1 - K^{\rm s})^{-1} A[y] B(y) \, dy 
$$
Analogous to the notation of \cite{TW94a} we put
$$
\phi^{\rm s}(x) := A(x) = {\rm Ai}(x), \qquad
Q^{\rm s}(x) := (1 - K^{\rm s})^{-1} A[x]
$$
so that
\begin{equation}\label{ue}
 \int_s^\infty (1 - K^{\rm s})^{-1} A[y] B(y) \, dy
= \int_s^\infty dy \, Q^{\rm s}(y) \int_{-\infty}^y 
\phi^{\rm s}(v) \, dv =: u_\epsilon^{\rm s}
\end{equation}
(the notation $u_\epsilon^{\rm s}$ -- without the
superscript s -- is used for an analogous quantity in
\cite{TW96}). Note from (\ref{8.stt1}) that with the notation
(\ref{ue}) we have
\begin{equation}\label{uem}
\Big ( E_1^{\rm s}(0;(s,\infty) \Big )^2 =
E_2^{\rm s}(0;(0;(s,\infty)) (1 - u_\epsilon^{\rm s}).
\end{equation}

Following \cite{TW96}, our objective is to derive coupled
differential equations for $u_\epsilon$ and the quantity
\begin{equation}\label{ue0}
q_\epsilon^{\rm s}
 := \int_s^\infty dy \, \rho^{\rm s}(s,y) 
\int_{-\infty}^y \phi^{\rm s}(v) \, dv
\end{equation}
where $\rho^{\rm s}(x,y)$ is the kernel of the operator 
$ (1 - K^{\rm s})^{-1}$. These equations will involve
\begin{equation}\label{ue1}
Q^{\rm s}(s) =: q_{\rm s} =  \int_s^\infty dy \, 
\rho^{\rm s}(s,y) \phi^{\rm s}(y),
\end{equation}
which in \cite{TW94a} is shown to be the Painlev\'e II transcendent
specified by the solution of (\ref{2.53}),
and their derivation relies on the formula \cite{TW94a}
\begin{equation}\label{ue2}
{\partial \over \partial s} Q^{\rm s}(y) = - q_{\rm s} \Big (
\delta^+(y-s) + \rho^{\rm s}(s,y) \Big ),
\end{equation}
where $\delta^+(y-s)$ is such that 
$$
\int_s^\infty \delta^+(y-s)f(y) \, dy = f(s),
$$
as well as the formula  \cite{TW94a}
\begin{equation}\label{ue3}
\Big ( {\partial \over \partial s} + {\partial \over \partial x} +
{\partial \over \partial y} \Big ) 
\rho^{\rm s}(x,y) = - Q^{\rm s}(x) Q^{\rm s}(y).
\end{equation}

Now, differentiating (\ref{ue}) with respect to $s$ we have
\begin{equation}\label{ue4}
(u_\epsilon^{\rm s})' = - q_{\rm s} \int_{-\infty}^s \phi^{\rm s}(v) \, dv +
\int_s^\infty dy \Big ( {\partial \over \partial s} Q^{\rm s}(y) \Big )
\int_{-\infty}^y \phi^{\rm s}(v) \, dv = - q_{\rm s} 
q_\epsilon^{\rm s}
\end{equation}
where to obtain the second equality use has been make of (\ref{ue2}) and
the definition (\ref{ue0}).
We now seek a formula for $(q_\epsilon^{\rm s})'$. Making use of
(\ref{ue3}) in (\ref{ue0}) shows
\begin{eqnarray}\label{ue5}
(q_\epsilon^{\rm s})' & = &
- \int_s^\infty dy \, {\partial \over \partial y} \rho^{\rm s}(s,y)
\int_{-\infty}^y \phi^{\rm s}(v) \, dv - q_{\rm s}
\int_s^\infty dy \, Q^{\rm s}(y) \int_{-\infty}^y \phi^{\rm s}(v) \, dv
\nonumber \\
& = &  \int_s^\infty  \rho^{\rm s}(s,y)  \phi^{\rm s}(y) \, dy
-  q_{\rm s} u_\epsilon^{\rm s} \nonumber \\
& = & q_{\rm s} (1 - u_\epsilon^{\rm s} )
\end{eqnarray}
where the final equality follows from the definitions (\ref{ue}) and
(\ref{ue1}).

As $ q_{\rm s}$ is known, the system of equations (\ref{ue4})
and (\ref{ue5}) fully determines $ u_\epsilon^{\rm s}$ and
$q_\epsilon^{\rm s}$ once boundary conditions are specified. Now
$Q^{\rm s}(y)$ is smooth, so we see from (\ref{ue}) that
\begin{equation}\label{d1}
u_\epsilon^{\rm s} \sim 0
\end{equation} 
as $s \to \infty$. On the other hand $\rho(s,y) = \delta^+(s-y)
+ R(s,y)$ where $R(s,y)$ is smooth, so for $s \to \infty$
\begin{equation}\label{d2}
q_\epsilon^{\rm s} \sim \int_{-\infty}^\infty  \phi^{\rm s}(v) \, dv
=  \int_{-\infty}^\infty {\rm Ai}(v)  \, dv = 1.
\end{equation}
The unique solution of the coupled
equations (\ref{ue4}) and (\ref{ue5}) satisfying (\ref{d1}) and
(\ref{d2}) is easily shown to be
\begin{equation}\label{d3}
u_\epsilon^{\rm s} = 1 - e^{-\mu_{\rm s}}, \qquad
q_\epsilon^{\rm s} =   e^{-\mu_{\rm s}}
\end{equation}
where 
$$
\mu_{\rm s} := \int_s^\infty q_{\rm s}(x) \, dx.
$$
Substituting the evaluation of $u_\epsilon^{\rm s}$ from
(\ref{d3}) in (\ref{uem}) reclaims (\ref{TW1}), as desired. 

\subsection{The probability $E_1^{\rm h}(0;(0,s);a)$}
All the steps leading to the rederivation of (\ref{TW1}) given in the
previous section have analogues for the probability
$E_1^{\rm h}(0;(0,s);a)$ which lead to the evaluation (\ref{R1}).

First, the analogue of (\ref{A1}) is
$$
\Big ( E_1^{\rm h}(0;(0,s);a) \Big )^2 =
E^{({\rm LOE})^2 {\rm h}}(0;(0,s);a),
$$
while use of the determinant formula (\ref{he10}) in (\ref{2a}) then gives
\begin{equation}\label{wf}
\Big ( E_1^{\rm h}(0;(0,s);(a-1)/2) \Big )^2 =
\det \Big ( 1 - (K^{\rm h} + C \otimes D) \Big ).
\end{equation}
Here $K^{\rm h}$ is the integral operator on $(0,s)$ with kernel
(\ref{KH}), while $C$ is the operator which multiplies by
$J_a(\sqrt{y})$, while $D$ is the integral operator with kernel
\begin{equation}\label{wf1}
{1 \over 2 \sqrt{y}} \int_{\sqrt{y}}^\infty J_a(t) \, dt.
\end{equation}
Recalling \cite{TW94b}
$$
\det (  1 - K^{\rm h} ) = E_2^{\rm h}(0;(0,s);a)
$$
we see that analogous to (\ref{8.stt1}), (\ref{wf}) can be rewritten
\begin{equation}\label{wf2}
\Big ( E_1^{\rm h}(0;(0,s);(a-1)/2) \Big )^2 =
 E_2^{\rm h}(0;(0,s);a) \Big (
1 - \int_0^s  (  1 - K^{\rm h} )^{-1} C[y] D(y) \, dy \Big ).
\end{equation}

Analogous to the notation of \cite{TW94b} we put
$$
\phi^{\rm h}(x) := C(x) = J_a(\sqrt{x}), \qquad
Q^{\rm h}(y) := (  1 - K^{\rm h} )^{-1} C[y].
$$
After changing variables $t = \sqrt{u}$ in (\ref{wf1}) we see that in terms
of this notation
\begin{equation}\label{wf3}
\int_0^s (  1 - K^{\rm h} )^{-1} C[y] D(y) \, dy =
{1 \over 4} \int_0^s dy \, Q^{\rm h}(y) {1 \over \sqrt{y}}
\int_y^\infty du \, {1 \over \sqrt{u}} \phi^{\rm h}(u) 
=: u_\epsilon^{\rm h},
\end{equation}
and in turn this latter notation used in (\ref{wf2}) gives
\begin{equation}\label{p1a}
\Big ( E_1^{\rm h}(0;(0,s);(a-1)/2) \Big )^2 =
 E_1^{\rm h}(0;(0,s);a) (1 - u_\epsilon^{\rm h}).
\end{equation}

Now, with 
$$
Q^{\rm h}(s) =: q_{\rm h} = \int_0^s dy \, \rho^{\rm h}(s,y) \phi^{\rm h}
(y),
$$
which in \cite{TW94b} is shown to be the Painlev\'e V transcendent specified
by the nonlinear equation (\ref{PV}),
analogous to (\ref{ue2}) we have \cite{TW94b}
$$
{\partial \over \partial s} Q^{\rm h}(y) =
q_{\rm h} \Big ( \delta^+(y-s) +  \rho^{\rm h}(s,y) \Big ).
$$
Use of this formula in (\ref{wf3}) then shows
\begin{equation}\label{p2}
(u_\epsilon^{\rm h})' = {1 \over 4} q_{\rm h} q_\epsilon^{\rm h}, \qquad
q_\epsilon^{\rm h} := \int_0^s dy \,
 \rho^{\rm h}(s,y) {1 \over \sqrt{y}} \int_y^\infty du \,
{1 \over \sqrt{u}} \phi(u)
\end{equation}
which is the analogue of (\ref{ue4}).

Next we seek a formula for the derivative with respect to $s$ of
$q_\epsilon^{\rm h}$. For this purpose we note from \cite{TW94b}
that
$$
x {\partial \over \partial x} 
\rho^{\rm h}(x,y) + s { \partial \over  \partial s}  \rho^{\rm h}(x,y)
= - { \partial \over  \partial y} \Big ( y  \rho^{\rm h}(x,y) \Big ) 
+ {1 \over 4} Q^{\rm h}(x) Q^{\rm h}(y)
$$
(c.f.~(\ref{ue3})). This formula applied to (\ref{p2}) shows
\begin{eqnarray}\label{p3}
s(q_\epsilon^{\rm h})' & = &
- \int_0^s dy \, \Big ( {d \over dy}(y \rho^{\rm h}(s,y) \Big )
{1 \over \sqrt{y}} \int_y^\infty du \, {1 \over \sqrt{u}}
\phi^{\rm h}(u) + q_{\rm h} u_\epsilon^{\rm h} \nonumber \\
& = & - {1 \over 2} \int_0^s dy \, \rho^{\rm h}(s,y) {1 \over \sqrt{y}}
\int_y^\infty du \, {1 \over \sqrt{u}} \phi^{\rm h}(u) -
\int_0^sdy \,  \rho^{\rm h}(s,y)  \phi^{\rm h}(y) +  q_{\rm h} 
u_\epsilon^{\rm h} \nonumber \\
& = & - {1 \over 2} q_\epsilon^{\rm h} - q_{\rm h} (1 - 
u_\epsilon^{\rm h})
\end{eqnarray}
The coupled equations (\ref{p2}) and (\ref{p3}) must be solved subject
to the $s \to 0$ boundary conditions
\begin{equation}\label{p4}
u_\epsilon^{\rm h} \sim 0, \quad
\sqrt{s} q_\epsilon^{\rm h} \sim \int_0^\infty {1 \over \sqrt{u}}
\phi(u) \, du = 2 \int_0^\infty J_a(v) \, dv = 2.
\end{equation}

The occurence of $\sqrt{s} q_\epsilon^{\rm h}$ in (\ref{p4}) suggests
we introduce
$$
\tilde{q}_\epsilon^{\rm h} := \sqrt{s} q_\epsilon^{\rm h}
$$
in (\ref{p2}) and (\ref{p3}). Doing this gives the system of equations
\begin{equation}\label{p5}
\sqrt{s} (u_\epsilon^{\rm h})' = {1 \over 4} q_{\rm h}
\tilde{q}^{\rm h}_\epsilon, \qquad
\sqrt{s}(\tilde{q}_\epsilon^{\rm h})' = - q_{\rm h}(1 -
u_\epsilon^{\rm h}).
\end{equation}
Introducing the new independent variable
$$
\mu_{\rm h} := \int_0^s {1 \over x^{1/2}} q_{\rm h}(x) \, dx
$$
we see that (\ref{p5}) reduces to the system with constant coefficients
\begin{equation}\label{p6}
{d \over d\mu} u_\epsilon^{\rm h} = {1 \over 4} \tilde{q}_\epsilon^{\rm h},
\qquad {d \over d\mu}  \tilde{q}_\epsilon^{\rm h} = - (1 -
u_\epsilon^{\rm h}).
\end{equation}
The solution satisfying (\ref{p4}) is
\begin{equation}\label{p7}
u_\epsilon^{\rm h} = 1 - e^{-{1 \over 2} \mu_{\rm h}}, \qquad
 \tilde{q}_\epsilon^{\rm h} = 2  e^{-{1 \over 2} \mu_{\rm h}}.
\end{equation}
The stated formula (\ref{R1}) for $E_1^{\rm h}(0;(0,s);(a-1)/2)$ now
follows by substituting this evaluation of $u_\epsilon^{\rm h}$ in 
(\ref{p1a}).

\section{Discussion}
\setcounter{equation}{0}
\subsection{Special values of $a$}
Edelman \cite{Ed88} was the first person to obtain the exact evaluation
of $E_1^{\rm h}(0;(0,s);a)$, albeit for two special values of $a$ only,
namely $a=-1/2$ and $a=0$. In terms of the scaling (\ref{1.SLa})
the results of  \cite{Ed88} are
\begin{eqnarray}
E_1^{\rm h}(0;(0,s); - {1 \over 2}) & = & e^{-(s/8 + \sqrt{s}/2)}
\label{E1} \\
E_1^{\rm h}(0;(0,s);0)  & = & e^{-s/8}. \label{E2}
\end{eqnarray}
Subsequently it was shown by the present author \cite{Fo93c} that 
$E_1^{\rm h}(0;(0,s);a)$ for $a \in \zz_{\ge 0}$ can be expressed as
a $2a$-dimensional integral. Explicitly
\begin{equation}\label{E3}
E_1^{\rm h}(0;(0,s);a) = C e^{-s/8} \Big ( {1 \over 2 \pi s^{1/2}}
\Big )^{2a} \int_{[-\pi, \pi]^{2a}}
\prod_{j=1}^{2a} e^{s^{1/2} \cos \theta_j} e^{i \theta_j}
\prod_{1 \le j < k \le 2a} | e^{i\theta_k} - e^{i \theta_j} |^4
d\theta_1 \cdots d\theta_{2a}
\end{equation}
where
$$
C = \prod_{j=1}^{2a} {\Gamma (3/2) \Gamma (3/2 + j/2) \over
\Gamma (1 + j/2) }.
$$
We remark that well known integration procedures (see e.g.~\cite{TW98}) allow
this integral to be expressed as a Pfaffian. Such Pfaffian formulas, deduced
in a different way, have been given in \cite{NF97i}.

The formula (\ref{R1}) relates $E_1^{\rm h}(0;(0,s);(a-1)/2)$ to
$E_2^{\rm h}(0;(0,s);a)$, so it is appropriate to consider the
evaluation of the latter for special values of $a$. Analogous to
(\ref{E3}) we have that for $a \in \zz_{\ge 0}$ \cite{Fo93c}
\begin{equation}\label{4.3'}
E_2^{\rm h}(0;(0,s);a) = e^{-s/4} \Big ( {1 \over 2 \pi} \Big )^a
{1 \over a!} \int_{[-\pi, \pi]^a}
\prod_{j=1}^a e^{s^{1/2} \cos \theta_j} \prod_{1 \le j < k \le a}
| e^{i \theta_k} - e^{i \theta_j} |^2 d\theta_1 \cdots d \theta_a
\end{equation}
This integral can easily be written as a Toeplitz determinant, with
the result \cite{FH94}
\begin{equation}\label{U1}
E_2^{\rm h}(0;(0,s);a) = e^{-s/4} \det \Big [ I_{j-k}(s^{1/2})
\Big ]_{j,k=1,\dots,a}
\end{equation}
where $I_n(x)$ denotes the Bessel function of purely imaginary argument.
As an aside, it is interesting to note that the integral (\ref{4.3'})
is the generating function for the enumeration of various combinatorial
objects, including quantities related to random permutations \cite{Ra98},
random words \cite{TW99} and random walks \cite{Fo99}.

The formula (\ref{FFa}) gives
\begin{equation}\label{U2}
- 4 {d \over ds} \Big ( s {d \over ds} \log
E_2^{\rm h}(0;(0,s);a) \Big ) = \Big ( q_{\rm h}(s) \Big )^2,
\end{equation}
so (\ref{U1})
implies that for $a \in \zz_{\ge 0}$, $ q_{\rm h}^2$ can be expressed
in terms of Bessel functions. The simplest case is $a=0$, when we have
\begin{equation}\label{E0}
E_2^{\rm h}(0;(0,s);a) \Big ) =   e^{-s/4}.
\end{equation}
Substituting in (\ref{U2}) gives \cite{TW94b}
\begin{equation}\label{E0'}
 q_{\rm h}(s) = 1
\end{equation}
and substituting this in  (\ref{R1}) we reclaim (\ref{E1}). The next
simplest case is $a=1$ when we have
\begin{equation}\label{ES}
E_2^{\rm h}(0;(0,s);1) =  e^{-s/4} I_0(s^{1/2})
\end{equation}
and so
$$
 \Big ( q_{\rm h}(s) \Big )^2 = 1 - 4
 {d \over ds} \Big ( s {d \over ds} \log  I_0(s^{1/2}) \Big ).
$$
For this to be consistent with  (\ref{R1}) we see that the
identity
$$
I_0(s^{1/2}) =
\exp \Big ( {1 \over 2} \int_0^s {1 \over \sqrt{t}}
\Big (  1 - 4
 {d \over dt} \Big ( t {d \over dt} \log  I_0(t^{1/2})
\Big ) \Big )^{1/2} \, dt
$$
must hold. This in turn is equivalent to the statement that
$y := \log I_0(s^{1/2})$ satisfies the nonlinear equation
$$
4s y'' + 4s (y')^2 +4 y' - 1 = 0,
$$
a fact which is readily verified using Bessel function identities.

Special evaluations are also known for $E_4^{\rm h}(0;(0,s);a)$
in the case $a \in \zz_{\ge 0}$ \cite{Fo93c}. 
These evaluations are in terms of a certain generalized hypergeometric
function based on Jack polynomials, while in the case $a$ even
Pfaffian formulas are also known \cite{NF97i}.
In quoting from
these results, one must be aware that $E_4^{\rm h}$ is defined
starting with the ensemble LSE${}_{N/2}$, and scaling according
to (\ref{1.SLa}). This means that the results of
\cite{Fo93c} require some rescaling of $s$. Doing this, we note
from \cite{Fo93c} that the simplest cases are $a=0$ and $a=1$, when we
have
\begin{eqnarray}
E_4^{\rm h}(0;(0,s);0) & = & e^{-s/8} \label{ad1} \\
E_1^{\rm h}(0;(0,s);1) & = & e^{-s/8} {}_0 F_1 \Big (
{1 \over 2}; - {s \over 16} \Big ) \: = \: e^{-s/8}
\cosh {\sqrt{s} \over 2}. \label{ad2}
\end{eqnarray}

The result (\ref{ad1}) can only be related to (\ref{R2}) in the limit
$a \to - 1^-$, since for $a \le -1$ $E_2^{\rm h} (0;(0,s);a) = 0$. However,
as the limiting forms of the quantities on the r.h.s.~are not known
we cannot readily check the consistency with (\ref{ad1}). In contrast
the consistency between (\ref{R2}) and (\ref{ad2}) is immediate upon
recalling (\ref{E0}) and (\ref{E0'}).

\subsection{Connection between $E_\beta^{\rm h}$ and 
$E_\beta^{\rm s}$}
In previous articles \cite{Fo93a,FNH99} it has been noted that for
$a \to \infty$, after appropriate rescaling of the coordinates, the
scaled $k$-point distribution function for the infinite Laguerre
ensemble at the hard edge becomes equal to the scaled
$k$-point distribution function for the infinite Gaussian ensemble
at the soft edge. Note that in the symplectic case the scalings are
done starting with the ensembles GSE${}_{N/2}$ and LSE${}_{N/2}$.
Let
$$
a(\beta) = \left \{ \begin{array}{ll} (a-1)/2, & \beta = 1 \\
a, & \beta = 2 \\
a+1, & \beta = 4 \end{array} \right.
$$
Explicitly, it was checked that the 
scaled $k$-point distribution function for the infinite Laguerre
ensemble at the hard edge,  with parameter $a \mapsto a(\beta)$
and after the rescaling of
 coordinates
$$
x \mapsto a^2 - 2a(a/2)^{1/3} x,
$$
equals
the soft edge distribution functions for the corresponding Gaussian ensemble
results in the $a \to \infty$ limit. We must therefore have
\begin{equation}\label{V1}
\lim_{a \to \infty} E_\beta^{\rm h}(0;(0,a^2 - 2a (a/2)^{1/3} s);
a(\beta) \Big ) = E_\beta^{\rm s}(0;(0,\infty)).
\end{equation}

To verify (\ref{V1}), we first recall some additional results from
\cite{TW94b}. Write
\begin{equation}\label{V8}
{1 \over 2} q_{\rm h} = \Big ( (s R_{\rm h})' \Big )^{1/2}
\end{equation}
so that, after integrating by parts, (\ref{FFa}) reads
\begin{equation}\label{V2}
E_2^{\rm h}(0;(0,s);a) = \exp \Big ( - \int_0^s R_{\rm h}(t) \, dt \Big ).
\end{equation}
Then $\sigma := s R_{\rm h}(s)$ is shown to satisfy the particular
Painlev\'e III equation in $\sigma$ form (for an account of the
latter see \cite{JM81,CS93})
\begin{equation}\label{QE1}
(s \sigma'')^2 + \sigma'(\sigma - s \sigma')(4\sigma'-1) - a^2
(\sigma')^2 = 0.
\end{equation}
Similarly let
\begin{equation}\label{V9}
q_{\rm s}(s) = (- R_{\rm s}'(s) )^{1/2}
\end{equation}
so that (\ref{3.53'}) reads
\begin{equation}\label{V3}
E_2^{\rm s}(0;(s,\infty)) = \exp \Big ( - \int_s^\infty R_{\rm s}(t)
\Big ).
\end{equation}
It is shown in \cite{TW94a} that $R_{\rm s}$ satisfies the particular
Painlev\'e II equation in $\sigma$ form
\begin{equation}\label{QE2}
(R_{\rm s}'')^2 + 4R_{\rm s}'\Big (
(R_{\rm s}')^2 - s R_{\rm s}' + R_{\rm s} \Big ) = 0.
\end{equation}
Substituting (\ref{V2}) and (\ref{V3}) in (\ref{V1}) with $\beta = 2$
we deduce that the validity of the latter is equivalent to the
statement
\begin{equation}\label{QE3}
2 a (a/2)^{1/3} R_{\rm h}(a^2 - 2a (a/2)^{1/3} t)
\mathop{\sim}\limits_{a \to \infty} R_{\rm s}(t).
\end{equation}
This can be verified by introducing the function
$$
\tilde{\sigma}(s) =
{2a(a/2)^{1/3} \over a^2} \sigma \Big ( a^2 - 2a(a/2)^{1/3} s
\Big ) \sim 2a(a/2)^{1/3}  R_{\rm h} \Big ( a^2 - 2a(a/2)^{1/3} s
\Big )
$$ 
into (\ref{QE1}) and taking the limit $a \to \infty$. One finds the
differential equation (\ref{QE2}) results with $R_{\rm s} =
\tilde{\sigma}(s)$.

In terms of (\ref{V8}), the evaluations (\ref{R1}) and (\ref{R2}) read
\begin{eqnarray*}
\Big ( E_1^{\rm h}(0;(0,s);(a-1)/2) \Big )^2 & = &
E_2^{\rm h}(0;(0,s);a) \exp \Big ( - \int_0^s
{((tR_{\rm h}(t))')^{1/2} \over \sqrt{t}} \, dt \Big ) \\
\Big ( E_4^{\rm h}(0;(0,s);a+1) \Big )^2 & = &
E_2^{\rm h}(0;(0,s);a)
\cosh^2 \Big ( {1 \over 2}
\int_0^s {(t R_{\rm h}(t))')^{1/2} \over \sqrt{t}} \, dt \Big ).
\end{eqnarray*}
Making use of (\ref{V1}) in the case $\beta = 2$ (which has just been
verified), and (\ref{QE3}) together with (\ref{V9}), we see that
\begin{eqnarray*}
\lim_{a \to \infty} \Big (  E_1^{\rm h}(0;(0,
a^2 - 2a(a/2)^{1/3}s);(a-1)/2) \Big )^2 & = &
E_2^{\rm s}(0;(s,\infty))
\exp \Big ( - \int_s^\infty q_{\rm s}(t) \, dt \Big ) \nonumber \\
\lim_{a \to \infty}
\Big ( E_4^{\rm h}(0;(0,a^2 - 2a(a/2)^{1/3}s);a+1) \Big )^2 & = &
E_2^{\rm s}(0;(s,\infty))
\cosh^2 \Big ( {1 \over 2} \int_s^\infty  q_{\rm s}(t) \, dt \Big ).
\end{eqnarray*} 
We recognize the right hand sides in these expressions as
$E_\beta^{\rm s}(0;(s,\infty))$ for $\beta = 1$ and 4 respectively
\cite{TW96} (recall eq.~(\ref{TW1}); as noted in \cite{FR99}
$E_4^{\rm s}(0;(s,\infty))$ can be deduced from 
$E_1^{\rm s}(0;(s,\infty))$ and $E_2^{\rm s}(0;(s,\infty))$ because
of the validity of an inter-relationship analogous to (\ref{1.15'})).

\subsection*{Acknowledgements}
This work was supported by the Australian Research Council.


\begin{thebibliography}{10}

\bibitem{AFNV99}
M.~Adler, P.J. Forrester, T.~Nagao, and P.~van Moerbeke.
\newblock Classical skew orthogonal polynomials and random matrices.
\newblock preprint, 1999.

\bibitem{AV99a}
M.~Adler and P.~van Moerbeke.
\newblock Symmetric random matrices and the {Pfaffian} {Toda} lattice.
\newblock solv-int/9903009, 1999.

\bibitem{BR99a}
J.~Baik and E.M. Rains.
\newblock Algebraic aspects of increasing subsequences.
\newblock math.CO/9905083, 1999.

\bibitem{Be97}
C.W.J. Beenakker.
\newblock Random-matrix theory of quantum transport.
\newblock {\em Rev. Mod. Phys.}, 69:731--808, 1997.

\bibitem{CS93}
C.M. Cosgrove and G.~Scoufis.
\newblock Painlev\'e classification of a class of differential equations of the
  second order and second degree.
\newblock {\em Stud. in Appl. Math.}, 88:25--87, 1993.

\bibitem{Dy70}
F.J. Dyson.
\newblock Correlations between the eigenvalues of a random matrix.
\newblock {\em Commun. Math. Phys.}, 19:235--250, 1970.

\bibitem{Ed88}
A.~Edelman.
\newblock Eigenvalues and condition numbers of random matrices.
\newblock {\em SIAM J. Matrix Anal. Appl.}, 9:543--560, 1988.

\bibitem{Fo93c}
P.J. Forrester.
\newblock Exact results and universal asymptotics in the {Laguerre} random
  matrix ensemble.
\newblock {\em J. Math. Phys.}, 35:2539--2551, 1993.

\bibitem{Fo93a}
P.J. Forrester.
\newblock The spectrum edge of random matrix ensembles.
\newblock {\em Nucl. Phys. B}, 402:709--728, 1993.

\bibitem{Fo99}
P.J. Forrester.
\newblock Random walks and random permutations.
\newblock math.CO/9907037, 1999.

\bibitem{FH94}
P.J. Forrester and T.D. Hughes.
\newblock Complex {Wishart} matrices and conductance in mesoscopic systems:
  exact results.
\newblock {\em J. Math. Phys.}, 35:6736--6747, 1994.

\bibitem{FNH99}
P.J. Forrester, T.~Nagao, and G.~Honner.
\newblock Correlations for the orthogonal-unitary and symplectic-unitary
  transitions at the hard and soft edges.
\newblock {\em Nucl. Phys. B}, 553:601--643, 1999.

\bibitem{FR99}
P.J. Forrester and E.M. Rains.
\newblock Inter-relationships between orthogonal, unitary and symplectic matrix
  ensembles.
\newblock solv-int/9907008, 1999.

\bibitem{Hs39}
P.L. Hsu.
\newblock On the distribution of the roots of certain determinantal equations.
\newblock {\em Ann. Eugen.}, 9:250--258, 1939.

\bibitem{JM81}
M.~Jimbo and T.~Miwa.
\newblock Monodromony preserving deformations of linear ordinary differential
  equations with rational coefficients {II}.
\newblock {\em Physica}, 2D:407--448, 1981.

\bibitem{Me71}
M.L. Mehta.
\newblock A note on correlations between eigenvalues of random matrices.
\newblock {\em Commun. Math. Phys.}, 20:245--250, 1971.

\bibitem{Mu82}
R.J. Muirhead.
\newblock {\em Aspects of multivariable statistical theory}.
\newblock Wiley, New York, 1982.

\bibitem{NF97i}
T.~Nagao and P.J. Forrester.
\newblock The smallest eigenvalue at the spectrum edge of random matrices.
\newblock {\em Nucl. Phys. B}, 509:561--598, 1998.

\bibitem{Ra98}
E.M. Rains.
\newblock Increasing subsequences and the classical groups.
\newblock {\em Elect. J. of Combinatorics}, 5:\#R12, 1998.

\bibitem{Sz75}
G.~Szeg\"o.
\newblock {\em Orthogonal polynomials}.
\newblock American Mathematical Society, Providence R.I., 4th edition, 1975.

\bibitem{TW94a}
C.A. Tracy and H.~Widom.
\newblock Level-spacing distributions and the {Airy} kernel.
\newblock {\em Commun. Math. Phys.}, 159:151--174, 1994.

\bibitem{TW94b}
C.A. Tracy and H.~Widom.
\newblock Level-spacing distributions and the {Bessel} kernel.
\newblock {\em Commun. Math. Phys.}, 161:289--309, 1994.

\bibitem{TW96}
C.A. Tracy and H.~Widom.
\newblock On orthogonal and symplectic matrix ensembles.
\newblock {\em Commun. Math. Phys.}, 177:727--754, 1996.

\bibitem{TW98}
C.A. Tracy and H.~Widom.
\newblock Correlation functions, cluster functions and spacing distributions in
  random matrices.
\newblock {\em J. Stat. Phys.}, 92:809--835, 1998.

\bibitem{TW99}
C.A Tracy and H.~Widom.
\newblock On the distributions of the longest monotone subsequences in random
  words.
\newblock math.CO/9904042, 1999.

\bibitem{Ve94}
J.~Verbaarschot.
\newblock The spectrum of the {Dirac} operator near zero virtuality for $n_c=2$
  and chiral random matrix theory.
\newblock {\em Nucl. Phys. B}, 426:559--574, 1994.

\bibitem{Wi28}
J.~Wishart.
\newblock The generalized product moment distribution in samples from a normal
  multivariate population.
\newblock {\em Biometrika}, 20A:32--43, 1928.

\end{thebibliography}

\end{document}